\documentclass[twocolumn,showpacs,preprintnumbers,amsmath,amssymb]{revtex4-1}

\usepackage{gensymb}
\usepackage{graphics}
\usepackage{graphicx}   
\usepackage{bm}
\usepackage[latin1]{inputenc}
\usepackage{textcomp}
\usepackage{float}

\begin{document}

\title{Photogating of mono- and few-layer MoS$_{2}$}

\author{Bastian Miller}	
	\affiliation{Walter Schottky Institut and Physik-Department, Technische Universit{\"a}t M{\"u}nchen, 85748 Garching, Germany}
          \affiliation{Nanosystems Initiative Munich (NIM), Schellingstr. 4, 80799 M{\"u}nchen, Germany }
\author{Eric Parzinger}
	\affiliation{Walter Schottky Institut and Physik-Department, Technische Universit{\"a}t M{\"u}nchen, 85748 Garching, Germany}
        \affiliation{Nanosystems Initiative Munich (NIM), Schellingstr. 4, 80799 M{\"u}nchen, Germany }
\author{Anna Vernickel}	
\affiliation{Walter Schottky Institut and Physik-Department, Technische Universit{\"a}t M{\"u}nchen, 85748 Garching, Germany}
         \affiliation{Nanosystems Initiative Munich (NIM), Schellingstr. 4, 80799 M{\"u}nchen, Germany }
\author{Alexander W. Holleitner}
\affiliation{Walter Schottky Institut and Physik-Department, Technische Universit{\"a}t M{\"u}nchen, 85748 Garching, Germany}
          \affiliation{Nanosystems Initiative Munich (NIM), Schellingstr. 4, 80799 M{\"u}nchen, Germany }
\author{Ursula Wurstbauer}
	\affiliation{Walter Schottky Institut and Physik-Department, Technische Universit{\"a}t M{\"u}nchen, 85748 Garching, Germany}
         \affiliation{Nanosystems Initiative Munich (NIM), Schellingstr. 4, 80799 M{\"u}nchen, Germany }
	 \email{wurstbauer@wsi.tum.de}
	
\date{\today}
\begin{abstract}
We describe a photogating effect in mono- and few-layer MoS$_{2}$, which allows the control of the charge carrier density by almost two orders of magnitude without electrical contacts. Our Raman studies are consistent with physisorbed environmental molecules, that effectively deplete the intrinsically n-doped charge carrier system via charge transfer, and which can be gradually removed by the exposure to light. This photogating process is reversible and precisely tunable by the light intensity. The photogating efficiency is quantified by comparison with measurements on electrostatically gated MoS$_{2}$.  
\end{abstract}


\maketitle
Two-dimensional layered van-der Waals materials and their heterostructures, particularly the transition metal dichalcogenides (TMDCs) family such as  MoS$_{2}$, are of great interest for fundamental research as well as for novel device concepts in the areas of  electronic \cite{Wang2012,Radisavljevic2011}, optoelectronic \cite{Mak2010,Yin2012,Wang2012,Lopez-Sanchez2013}, valley- and spintronic \cite{Xu2014} as well as solar energy conversion \cite{Tsai2014} and sensing applications \cite{Late2013}. It has been demonstrated that MoS$_{2}$ undergoes a transition from an indirect to a direct bandgap semiconductor by thinning it down to one single layer \cite{Mak2010}. The direct gap at the $K$-point with $E_{gap}$ = 1.9 eV \cite{Mak2010} remains almost unaffected by the number of layers. The indirect gap  existing in vicinity of the $\Gamma$-point decreases in energy for increasing number of layers \cite{Mak2010}. In particular, a high absorption of the  monolayer  of 5-10$\%$ in the visible regime \cite{Bernadi2013} makes MoS$_{2}$ a promising material for optoelectronic devices such as phototransistors and solar cells. For atomistic thin materials, knowledge about the interaction with the dielectric environment is of great importance. The impact of substrates  \cite{Li2013,Buscema2014}, the gaseous environment as well as of adsorbates \cite{Tongay2013,Mao2013,Nan2014} on the optoelectronic properties of 2D materials have been reported. The atomistic interfaces offer the opportunity of novel route for interfacial engineering of electronic, optical and optoelectronic properties \cite{Tongay2013,Yue2013,Li2013,Najmaei2014}.
\par
In this letter, we study the effect of photogating in MoS$_{2}$ mono- and few-layer flakes by means of Raman spectroscopy. Our observations are consistent with physisorbed environmental molecules acting as molecular gates. The molecules can be gradually removed from the MoS$_{2}$ surface by the exposure to light. We find that the photogating effect facilitates the control of the charge carrier density in MoS$_{2}$ by almost two orders of magnitude in pristine flakes without the need for electrical contacts. In general, Raman spectroscopy provides access to various properties of 2D layered materials \cite{Pimenta2015} such as to the number of layers \cite{Lee2010,Plechinger2012}, to strain \cite{Rice2013, Conley2013} to the local temperature \cite{Lanzillo2013} and to the doping level \cite{Chakraborty2012}. The Raman spectrum of MoS$_{2}$ includes two prominent first-order phonon modes, the E$^{1}_{2g}$ and the A$_{1g}$ mode \cite{Pimenta2015}. The E$^{1}_{2g}$ mode is an in-plane mode, i.e. the atoms are oscillating parallel to the basal plane of the van-der Waals coupled crystal layers. The A$_{1g}$ mode is an out-of plane oscillation, where the sulfur atoms are moving in opposite directions. Raman studies on mono- and few-layer MoS$_{2}$ reveal, that the energy of the E$^{1}_{2g}$ mode decreases with increasing number of layers, whereas the energy of the A$_{1g}$ mode increases with the number of layers \cite{Lee2010}. Moreover, it has been reported for a monolayer of MoS$_{2}$ that the energy of the A$_{1g}$ mode is sensitive to the doping level due to phonon renormalization \cite{Chakraborty2012}.
\par
\begin{figure}
\centering
\includegraphics[width=0.5\textwidth]{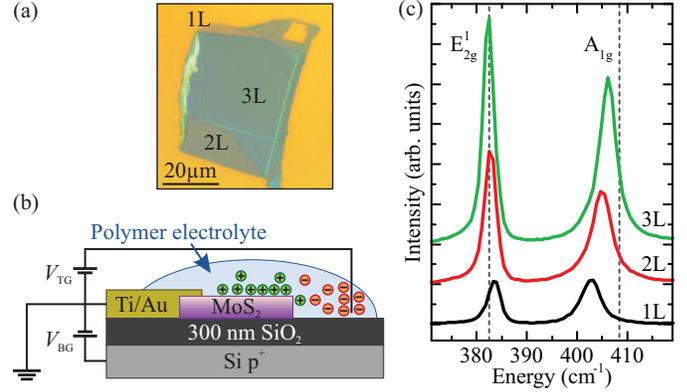}
\caption{(a) Optical microscopy image (100x) of a MoS$_{2}$ flake exibiting mono-, bi- and tri-layer areas. (b) Scheme of a top- and back-gated MoS$_{2}$ field-effect device for optical measurements. (c) Raman spectra of mono-, bi-, tri-layer areas of the MoS$_{2}$ flake shown in (a). The dashed lines mark the phonon energies for bulk MoS$_{2}$.}
\label{fig:fig1}
\end{figure}
\par
In our experiments, laser power dependent Raman spectra taken on pristine mono- and few-layers of MoS$_{2}$ on Si/SiO$_{2}$ and sapphire substrates in ambient conditions and vacuum are contrasted with Raman measurements taken on electrostatically doped MoS$_{2}$ utilizing a field-effect device on SiO$_{2}$ substrate. For mono- bi-, tri- and four-layer flakes, we find a laser power induced shift of the A$_{1g}$ phonon mode energy. 
From a comparison with measurements done on electrostatically gated mono- and bilayer MoS$_{2}$ devices, we interpret the light induced shift as a change in the charge carrier density $\Delta n$ via a photogating effect. We further find that for up to 4 layers, the photogating effect scales linearly to the inverse number of layers. Therefrom, we conclude that the free charge carriers are delocalized and homogeneously distributed over all layers for all investigated samples. The observed photogating effect is not persistent, not observable in vacuum and independent from the investigated substrates. The photogating effect is of importance for future mono- and few-layered MoS$_{2}$ optoelectronic devices operating in ambient conditions.
\par
The studied MoS$_{2}$ flakes are micromechanically exfoliated from naturally occurring bulk crystals. They are transferred on sapphire or Si/SiO$_{2}$ substrates [Fig.\ref{fig:fig1}(a)]. The latter consists of p$^{+}$-doped silicon with a 300 nm thick SiO$_{2}$ layer [Fig. \ref{fig:fig1}(b)]. All flakes are characterized by optical microscopy [Fig.\ref{fig:fig1} (a)] and $\mu$-Raman spectroscopy in back scattering geometry under ambient conditions at room temperature with $\lambda =$ 488 nm  [Fig. \ref{fig:fig1}(c)]. The power of the exciting laser is varied from 25 to 1000 $\mu$W resulting in a power density between 0.5 kW/cm$^{2}$ and 56.6 kW/cm$^{2}$. 
Ohmic contacts for the dual gated MoS$_{2}$ field-effect device [Fig. \ref{fig:fig1}(b)], are prepared by optical lithography and e-beam evaporation of Ti/Au (5 nm/20 nm). The doped Si substrate serves as a back gate electrode. A solid polymer electrolyte (PE) composed of a mixture of polyethylene-oxide and CsClO$_{4}$ act as a transparent top gate electrode \cite{Das2008}. The capacity of such an electrode is as high as $C_{TG}\approx 1.5$ $\mu$F/cm$^{2}$ ~\cite{Chakraborty2012} providing a charge injection parameter of $\alpha_{PE} \approx 0.9 \times 10^{13}$ cm$^{-2}/$V. The MoS$_{2}$ flakes are intrinsically n-type doped as confirmed from complementary electrical measurements on a field-effect transistor device (data not shown).
\par
The Raman spectra of the MoS$_{2}$ flake depicted in Fig. \ref{fig:fig1}(a) are shown for the mono-, bi- and tri-layer areas in Fig. \ref{fig:fig1}(c). The energies of the E$^{1}_{2g}$ and the A$_{1g}$ modes as well as their energy difference $\Delta E = |E_{A}-E_{E}|$ are in excellent agreement to reported values in literature \cite{Lee2010}. We note that for the field-effect device, the energy of the A$_{1g}$ phonon is red shifted by about 0.6 cm$^{-1}$ after deposition of the PE top gate. We ascribe this shift to a moderate modification of the charge carrier density in MoS$_{2}$ introduced from a different dielectric environment on the top interface by adding the PE. The E$^{1}_{2g}$ mode is unaffected by the presence of the PE electrode within the experimental error of $\pm 0.2$ cm$^{-1}$. This excludes strain to be introduced by the polymer film covering the MoS$_{2}$ flake \cite{Rice2013}.
\par
\begin{figure}[ht]
\centering
\includegraphics[width=0.48\textwidth]{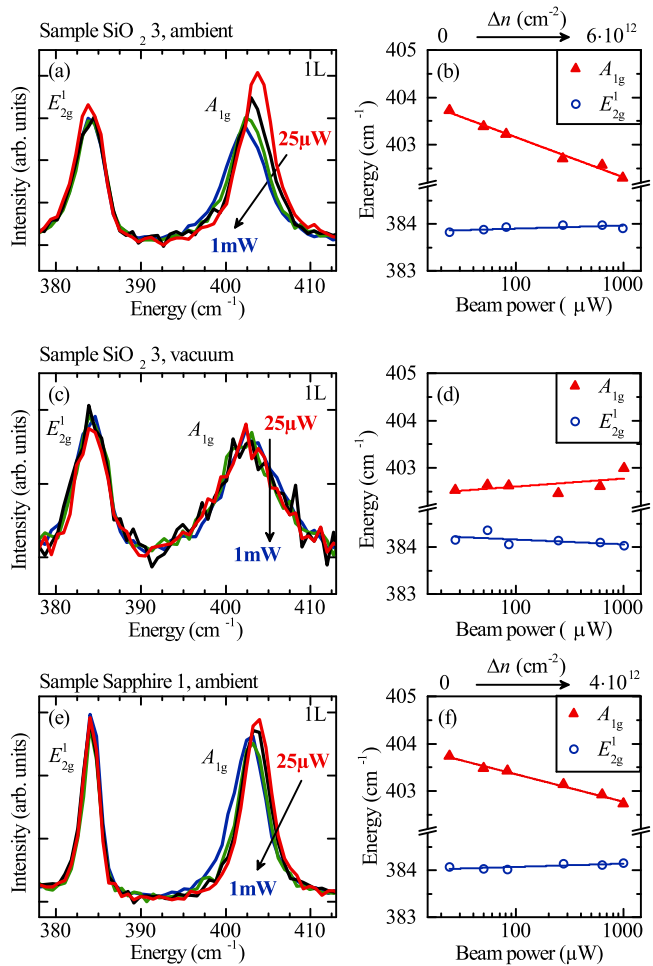}
\caption{Raman spectra of monolayer MoS$_{2}$ for different laser powers ranging from 25 to 1000 $\mu$W (a) on Si/SiO$_{2}$ taken in ambient conditions, (c) on Si/SiO$_{2}$ in vacuum and (e) on sapphire in ambient.  (b), (d), (f) Semi-logarithmic representation of the related A$_{1g}$ and the E$^{1}_{2g}$ mode energies in dependence of the laser power. The solid lines are linear fits to the data points.}
\label{fig:fig2}
\end{figure}
\par
Fig. \ref{fig:fig2} shows Raman spectra in dependence of the laser power of pristine monolayer MoS$_{2}$ on SiO$_{2}$ and on sapphire in ambient conditions as well as on SiO$_{2}$ in vacuum [Figs. \ref{fig:fig2}(a,c,e)]. For the measurements performed in ambient conditions [Figs. \ref{fig:fig2}(a,e)], the A$_{1g}$ mode shifts towards lower energies. Furthermore, it broadens with increasing laser power, while the E$^{1}_{2g}$ mode remains almost unaffected. Figs. \ref{fig:fig2}(b,d,f) show the energies of the A$_{1g}$ and E$^{1}_{2g}$ modes  vs. the laser power in a semi-logarithmic representation. Within the experimental resolution, an exponential dependence of the A$_{1g}$ mode energy from the laser power is observable for all samples measured in ambient conditions. The energy of the E$^{1}_{2g}$ phonon mode is unaffected by the laser power. Therefore, light-induced heating of the MoS$_{2}$ flake can be ruled out \cite{Lanzillo2013}. As discussed below, we attribute the redshift of the A$_{1g}$ mode to a phonon renormalization by an increase of free charge carriers; consistent with earlier reports on an electrostactically doped monolayer MoS$_{2}$ \citep{Chakraborty2012}.
\par
\begin{figure}[h]
\centering
\includegraphics[width=0.5\textwidth]{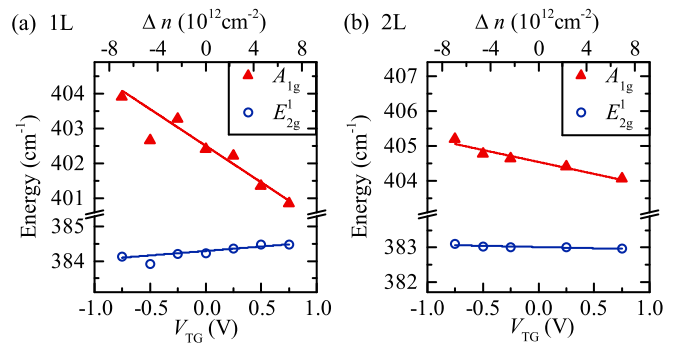}
\caption{Energy of the A$_{1g}$ and the E$^{1}_{2g}$ modes in dependence of the applied top gate voltage for mono-layer (a) and bilayer MoS$_{2}$ (b). The solid lines are linear fits to the data points.}
\label{fig:fig3}
\end{figure}

To corroborate our interpretation, we perform complementary Raman studies on electrostatically gated MoS$_{2}$ mono- and bi-layer field-effect devices.  Fig. \ref{fig:fig3} encloses that the A$_{1g}$ mode energy is linearly dependent on the applied gate voltage and hence on the change in charge carrier density $\Delta n$ for both MoS$_{2}$ mono- and bi-layer flakes. Again, the E$^{1}_{2g}$ mode is unaffected by the gate voltage within the experimental range. The gate induced change in the charge carrier density $\Delta n$ can be quantified from the charge injection parameter $\alpha$ [top axes in Fig. \ref{fig:fig3}]. Consequently, the light induced change $\Delta n$ can be deduced from a direct comparison with gate-voltage dependent measurements [top axes in Figs. \ref{fig:fig2}(b) and \ref{fig:fig2}(f)]. With both methods, the charge carrier density can be tuned by almost two orders of magnitude identified by a shift of the $A_{1g}$ phonon mode of $\approx$ 3 cm$^{-1}$ for mono- and $\approx$ 1.5 cm$^{-1}$ for bi-layers, respectively.
\par
\begin{table}[h]
\centering
\begin{tabular}{ccccc}\hline\hline
sample&$ E_{A}$&FWHM&$\Delta E_A$& $\Delta n$\\
(environment)&(cm$^{-1}$)&(cm$^{-1}$)& (cm$^{-1}$)& ($10^{12}$cm$^{-2}$)\\\hline
SiO$_{2}$ 1 (a)& 404.1& 3.6&-0.6&2.8\\
SiO$_{2}$ 1 (v)&402.6 & 5.9&0.0&$<$ 0.1\\
SiO$_{2}$ 2 (a)&404.3 & 3.3&-0.3&1.5\\
SiO$_{2}$ 3 (a)&403.7 & 4.1&-1.4&6.0 \\
SiO$_{2}$ 3 (v)&402.5 & 7.0&0.0&$<$0.0\\
SiO$_{2}$ 4 (a)&403.1 & 5.2&-2.3&10.0 \\
Sapphire 1 (a)& 403.7& 3.6&-0.9&4.1\\
Sapphire 2 (v)&402.1&6.9&$<$-0.1&$<$ 0.6\\
\hline\hline  
\end{tabular}
\caption{Energy $E_{A}$ and full width at half maximum (FWHM) of the A$_{1g}$ mode (P = 25 $\mu$W) measured in ambient (a) and vacuum (v) as a measure for the intrinsic charge carrier density. Photogating efficiency given by the decrease of the A$_{1g}$ mode energy  $\Delta E_{A}$ by increasing the power from P = 25 $\mu$W to P = 1 mW. $\Delta E_A$ has been translated into a change in the electron density $\Delta n$  by comparison with electrostatic doping.}
\label{tab:tab}
\end{table}
\par
We note that the described photogating effect is not persistent and it is independent from the used substrate. It is reversible and tunable by the power of the light. Photogating is not observable for measurements in vacuum as shown exemplarily for MoS$_{2}$ on SiO$_{2}$ in Figs. \ref{fig:fig2} (c,d). These findings strongly suggest that the interface between MoS$_{2}$ and the gaseous environment causes the laser power dependent change in the charge carrier density $\Delta n$. Additionally, the A$_{1g}$ mode is significantly broadened and red-shifted for measurements in vacuum with $p \leq 5 \times 10^{-6}$ mbar [Fig. \ref{fig:fig2} (c)] compared to measurements in ambient conditions. This behavior suggests a large charge carrier density of MoS$_{2}$ in vacuum, which is independent from the laser power [Fig. \ref{fig:fig2}(d)]. The enhanced doping in vacuum is entirely reversible. After measurements in vacuum and exposing the MoS$_{2}$ sheets to ambient air, the A$_{1g}$ mode is again stiffened and the FWHM is reduced. We further point out that the absolute values of the A$_{1g}$ mode energy and FWHM vary from sample to sample (compiled for 6 samples in table \ref{tab:tab}). However, the individual values are stable against repetition of measurements in vacuum and ambient conditions as well as measurements with varying light intensities.
\par
\par
We explain the photogating by a photodesorption of environmental molecules that are physisorbed on the  MoS$_{2}$ surface. It has been reported in literature from photoluminescence measurements that physisorption of H$_{2}$O and O$_{2}$ molecules leads to a reduction of the charge carrier density in MoS$_{2}$ \cite{Tongay2013,Nan2014}.The physisorbed molecules act as molecular gates \cite{Tongay2013, Yue2013} causing a charge transfer from the MoS$_{2}$ surface to the molecules of $\approx$ 0.01 electrons per H$_{2}$O molecule and $\approx$ 0.04 electrons per O$_{2}$ molecule \cite{Tongay2013, Nan2014}. These values are for defect free crystals. The charge transfer from MoS$_{2}$ to molecules that are physisorbed on defect sites or wrinkles might be altered. Removing such molecules prevents the charge transfer from the MoS$_{2}$ layers resulting in an increasing charge carrier density. The binding energies of physisorbed H$_{2}$O and O$_{2}$ molecules on the surface of MoS$_{2}$ determined from density functional theory (DFT) calculations are 110 meV and 79 meV, respectively \cite{Tongay2013,Nan2014}. Consistently, both molecules can be easily removed with measurements under high-vacuum conditions, especially during the exposure to light. 
\par
Let us now turn to the discussion of the laser power dependent increase of the charge carrier density for measurements in ambient conditions. We find a softening and broadening of the A$_{1g}$ phonon mode for measurements performed in vacuum as well as for measurements under ambient conditions with a high laser power of $P=1m$W. In line with the removal of a large amount of physisorbed molecules in vacuum, we interpret the laser power dependence of the A$_{1g}$ mode as a consequence of laser induced cleaning of the MoS$_{2}$ surface. In this picture, an increase of the laser power diminishes the number of physisorbed molecules and hence, the charge transfer from MoS$_{2}$ to the molecules is reduced. In short, for a low laser power a larger amount of physisorbed environmental molecules exists that effectively deplete the intrinsic electron density in MoS$_{2}$. Vice versa, a high laser power diminishes the amount of physisorbed molecules (laser cleaning) resulting in a reduction of the charge transfer from MoS$_{2}$ and a large electron density within the 2D crystal.  
\par
\begin{figure}[htpb]
\centering
\includegraphics[width=0.48\textwidth]{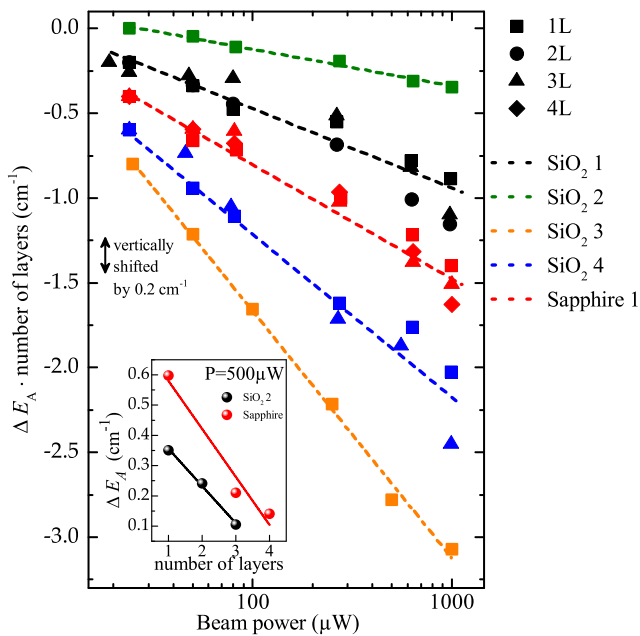}
\caption{Laser power dependent energy shift $\Delta E_{A}$ of the A$_{1g}$ phonon mode (relative to the $A_{1g}$ mode energy at $P = 25 \mu$W) for different samples and number of layers in ambient conditions. The deduced energy shift $\Delta E_{A}$ is multiplied by the number of layers. The traces are shifted vertically by $0.2$ cm$^{-1}$ for clarity. Inset: $\Delta E_{A}$ vs number of layers for to sample at an laser power of $P$ = 500 $\mu$W.}
\label{fig:fig4}
\end{figure}
In a steady state illumination the amount of physisorbed molecules is given by laser cleaning and physisorption rates. Whereas the physisorption rate is fixed for a specific sample and its environment, the laser cleaning rate depends on the light intensity. In this way, we  explain the tunability of the laser power induced change in the charge carrier density.
Differences between the individual samples are expected to be caused by crystal defects such as sulfur vacancies in the MoS$_{2}$ lattice\cite{Nan2014} or ripples acting as attractive sites for physisorption of environmental molecules. The used substrate materials seem to play an inferior role compared to the statistical variation from flake to flake exfoliated from the identical bulk crystal. 
\par
Fig. \ref{fig:fig4} shows the laser power dependent shift of the A$_{1g}$ phonon mode energy $\Delta E_{A}$ relative to the mode energy at $P$ = 25$\mu$W for different flakes consisting of mono- and few-layer parts. Intriguingly, we find that $\Delta E_{A}$ scales linearly with the inverse number of layers. In Fig. \ref{fig:fig4}, the shift of the A$_{1}$ mode energy multiplied by the number of layers ($\Delta E_{A} \times$ number of layers) is plotted vs. the laser power for 6  samples. The scaling with respect to the number of layers [inst Fig. \ref{fig:fig4}] reveals that the phonon-renormalization has a similar influence on the A$_{1g}$ mode energy for mono-, bi-, tri- and four-layer MoS$_{2}$ under the assumption that free charge carriers are delocalized and freely distributed over all individual basal planes. The same proportionality is observable for electrostatically induced changes in the electron density for MoS$_{2}$ mono- and bi-layer flakes [Fig. \ref{fig:fig3}].
\par
In summary, we show that the charge carrier density in MoS$_{2}$-devices in ambient conditions is precisely tunable by the light intensity.  This photogating effect is studied by Raman spectroscopy by the sensitivity of the A$_{1g}$ phonon mode to the charge carrier density. The photogating is quantified from comparison with gate-voltage dependent measurements on field effect devices. We further conclude that the charge carrier density is homogeneously distributed over the basal planes for up to 4 layers. Our study reveals that the photogating can be assigned to the light induced removal of physisorbed molecules acting as molecular gates. This effect facilitates to reversibly modify the interfacial properties of optoelectronic MoS$_{2}$ devices  without the need for electrical contacts. 
\par
We acknowledge fruitful discussions with Joel W. Ager and financial support by the DFG via excellence cluster Nanosystems Initiative Munich (NIM) and project Ho 3324/8-1 as well as BaCaTec.

%

\

\end{document}